\documentclass[12pt]{article}
\usepackage{amssymb}
\usepackage{amsmath, amsfonts}
\usepackage{latexsym}
\usepackage[all]{xy}
\input xy
\xyoption{all}
\begin{document}
\def\mkb{\mbox}
\def\beq{\begin{equation}}
\def\eeq{\end{equation}}
\def\beqn{\begin{eqnarray}}
\def\eeqn{\end{eqnarray}}
\def\H{\mathcal{H}}
\def\F{\mathcal{F}}
\def\P{\mathcal{P}}
\def\A{\mathcal{A}}
\def\l{\mathfrak{l}}
\def\L{\mathcal{L}}
\def\U{\mathcal{U}}
\def\T{\mathcal{T}}
\def\aT{{\bar{\T}(\A)}}
\def\X{\mathcal{X}}
\def\g{\mathfrak{g}}
\def\ta{{\tilde{\phi}}}
\def\bd{\boxdot}
\def\bp{\boxtimes}
\def\B{B_e^+}
\def\C{\mathbb{C}}
\def\R{\mathbb{R}}
\def\K{\mathbb{K}}
\def\N{\mathbb{N}}
\def\Z{\mathbb{Z}}
%
\thispagestyle{empty}
$\phantom{d}$
\vspace{3cm}
\begin{center}
{\LARGE{On the associative Nijenhuis Relation}}\\[0.1cm]
\end{center}
\vspace{2cm}
\begin{center}
%
%
%
%
    K. Ebrahimi-Fard
\footnote{
          e-mail: fard@th.physik.uni-bonn.de, kurro@math.bu.edu\\
          on leave from Bonn University, Phys. Inst., Theory Dept.
         }\\
\vspace{2cm}
{\sl Center for Mathematical Physics}\\
{\sl at Boston University}\\
{\sl 111 Cummington Street,}\\
{\sl Boston, MA 02215, USA}\\
\end{center}
\vspace{4cm} \setcounter{page}{1}
\begin{abstract}
In this brief note we would like to give the construction of a
free commutative unital associative Nijenhuis algebra on a
commutative unital associative algebra based on an augmented
modified quasi-shuffle product.\\
\end{abstract}
---------------------------------------\\
{\tiny{
\begin{tabular}{ll}
Keywords: & quasi-shuffle, modified quasi-shuffle, associative Nijenhuis relation,\\
          & Rota-Baxter relation\\
MSC-class:& 08B20, 83C47, 35Q15, 17A30\\
\end{tabular}        }}
\newpage
%
%
\section{Introduction}
%
The associative analog of the Nijenhuis relation \cite{CGM} may be
regarded as the homogeneous version of the Rota-Baxter relation
\cite{Rota1, Rota2, Rota3, Rota4, A, B, C, KEF}. Some of its
algebraic aspects especially with regard to the notion of quantum
bihamiltonian systems were investigated by Cari{\~n}ena {\emph{et
al.}} in \cite{CGM}. The Lie algebraic version of the associative
Nijenhuis relation is investigated in \cite{GS1, GS2} in the
context of the classical Yang-Baxter equation which is closely
related to the Lie algebraic version of the Rota-Baxter relation
(see especially \cite{BD}). Likewise in \cite{KSM, KS} the
deformation of Lie brackets defined by Nijenhuis operators and the
connection to the (modified) classical Yang-Baxter relation are
studied respectively. In Kreimer's work \cite{Kreimer1, Kreimer2}
the connection of the Rota-Baxter relation to the Riemann-Hilbert
problem in the realm of perturbative Quantum Field Theory is reviewed.\\
The algebraic properties of the associative notion of the
Nijenhuis relation, respectively Nijenhuis algebras, provide
interesting insights into associative analogs of Lie algebraic
structures.
Giving the definition of a commutative unital associative
Nijenhuis algebra we use an augmented modified quasi-shuffle
product to give explicitly the construction of the free
commutative unital associative Nijenhuis algebra generated by a
commutative unital associative $\K$-algebra. We follow thereby
closely the inspiring work of Guo {\emph{et al.}} \cite{LG1, LG2}.
The main aspect of this construction of the free object is the use
of a ''homogenized'' notion of Hoffman's quasi-shuffle product
giving an associative, unital, and commutative composition. This
natural ansatz relies on the close resemblance between the
Rota-Baxter and the associative Nijenhuis relation. The free
construction of the former is essentially given by the
quasi-shuffle product.\\
Let us remark here that in recent papers \cite{KEF, Agu1, Agu2} it
was shown that non-commutative\footnote{not necessarily
commutative} Rota-Baxter algebras always give Loday-type algebras,
i.e. dendriform di- and trialgebra structures \cite{JLL1, JLLR,
JLL3}.\\ 
The paper is organized as follows. In the first section we give
the definition of a commutative, unital and associative Nijenhuis
$\K$-algebra respectively the associative Nijenhuis relation. We
introduce the notion of a Nijenhuis homomorphism and a free
Nijenhuis algebra. Section two contains the definition of the
modified and augmented modified quasi-shuffle product being
commutative, having a unit and of which we prove the associativity
property explicitly. In the following section the augmented
modified quasi-shuffle product algebra is identified as the free
commutative associative unital Nijenhuis algebra, thereby giving
an explicit construction of it. This section closes with a couple
of remarks especially concerning the relation to Loday-type
algebras. We end this work with a short summary and outlook in the
final section.\\[0.2cm]
Throughout this paper, we will consider $\K$ to be a commutative
field of characteristic zero. The term $\K$-algebra always means
if not stated otherwise associative commutative unital
$\K$-algebra.
%
%
%
%
%
\section{The associative Nijenhuis Relation}
%
Let $\A$ be a $\K$-algebra (see last remark above). On $\A$ we
have the $\K$-linear map $N: \A \to \A$. We call $\A$ a Nijenhuis
$\K$-algebra in case the operator $N$ holds the following so
called associative Nijenhuis relation \cite{CGM}:
\beq
    N(x)N(y) + N^2(xy) = N\big(N(x)y + xN(y)\big), \;\; x,y \in \A.
    \label{ANR}
\eeq
The map $N$ might be called associative Nijenhuis operator or just
Nijenhuis map for short. In this setting ''associative'' refers to
the relation (\ref{ANR}) to distinguish it clearly form its Lie
algebraic version \cite{CGM, GS1, GS2, KSM, KS}:
\beq
    [N(x),N(y)] + N^2([x,y]) = N\big([N(x),y] + [x,N(y)]\big).
    \label{LNR}
\eeq
A Nijenhuis map on a $\K$-algebra $\A$ gives also a Nijenhuis map
for the associated Lie algebra $(\A,[\;,\;])$, $[\;,\;]$ being the
commutator.\\
Equation (\ref{ANR}) may be interpreted as the homogeneous version
of the standard form of the Rota-Baxter relation of weight
$\lambda$ \cite{KEF}:
\beq
    R(x)R(y) + \lambda R(xy) = R\big(R(x)y + xR(y)\big)
    \label{RBR}.
\eeq
A simple transformation $R \to \lambda^{-1}R$ gives the standard
form of (\ref{RBR}). The homogeneity of (\ref{ANR}) destroys this
freedom to renormalize the operator $N$ so as to allow for either
sign in front of the second term on the lefthand side of
(\ref{ANR}) for instance.\\
Let us give two examples of operators fulfilling the associative
Nijenhuis relation. Left or right multiplication $L_{a}b:=ab$,
$R_{a}b:=ba$, $a,b \in \A$ both hold relation (\ref{ANR}).\\
Another class of associative Nijenhuis operators comes from
idempotent Rota-Baxter operators. Let $R$ be such an idempotent
($R^2=R$) Rota-Baxter operator (\ref{RBR}). The operator
$\widehat{R}:=1-R$ being idempotent, too, also holds equation
(\ref{RBR}). Define the following operator:
\beq
    N_{\tau} := R - \tau \widehat{R}, \;\; \tau \in \K.
\eeq
The map $N_{\tau}$ is an associative Nijenhuis operator. One finds
similar examples and further algebraic aspects related to equation
(\ref{ANR}) in \cite{CGM}.  \\[0.4cm]
{\emph{Nijenhuis Homomorphism and the Universal Property}}

%
We call an algebra homomorphism $\phi: \A_1 \to \A_2$ between
Nijenhuis (Rota-Baxter) $\K$-algebras $\A_1$ respectively $\A_2$ a
Nijenhuis (Rota-Baxter) homomorphism if and only if $\phi$
intertwines with the Nijenhuis (Rota-Baxter) operators $N_1,\;
N_2$ ($R_1,\; R_2$):
\beq
    \phi \circ X_1 = X_2 \circ \phi,\;\; X_i= N_i\; (R_i),\; i=1,2.
    \label{NP}
\eeq
A Nijenhuis (Rota-Baxter) $\K$-algebra $F(\A)$ generated by a
$\K$-algebra $\A$ together with an associative algebra
homomorphism $j_{\A}: \A \to F(\A)$ is called a free Nijenhuis
(Rota-Baxter of weight $\lambda \in \K$) $\K$-algebra if the
following universal property holds: for any associative Nijenhuis
(Rota-Baxter) $\K$-algebra $\X$ and $\K$-algebra homomorphism
$\phi: \A \to \X$, there exists a unique Nijenhuis (Rota-Baxter)
homomorphism ${\tilde{\phi}}:
F(\A) \to \X$ such that the diagram:\\
%
\[
\xymatrix{
   \A    \ar[d]_{j_{\A}} \ar[rd]^{\phi} &\\
   F(\A) \ar[r]_{{\tilde{\phi}}} & \X                                 \\
   }
\]
commutes.\\
In the next section we introduce Hoffman's quasi-shuffle product
and an augmented respectively augmented modified version of it to
give the construction of the free Nijenhuis algebra.
%
%
%
%
%
\section{Modified Quasi-Shuffle Product}
%
Let $\A$ be a $\K$-algebra. We denote the product on $\A$ by $[ab]
\in \A$, $a,b \in \A$ and the algebra unit by $e \in \A$,
$[ea]=a,\; a \in \A$.\\
Consider the tensor module of $\A$:
\beq
    \T(\A) :=  \bigoplus_{n \ge 0} \A^{{\otimes}^n}.
\eeq
We denote generators \makebox{$a_1 \otimes \dots \otimes a_n$} by
concatenation, i.e. words \makebox{$a_1 \dots a_n$}. Generally we
use capitals \makebox{$U \in \A^{\otimes ^n},\; n>0$} for words
and lower case letters $a \in \A$ for letters. The natural grading
on ${\T}(\A)$ is defined by the length of words, \makebox{$\l(U +
V):= \l(U) + \l(V)= n+m$}, $U \in \A^{\otimes ^n}$, $V \in
\A^{\otimes ^m}$, $n,\: m > 1$, and $\l(a)=1$.
The empty word is denoted by $1$ and $\l(1)=0$.\\
We denote the associative commutative quasi-shuffle product
\cite{MH} on $\T(\A)$ by:
\beq
    aU*bV := a(U*bV) + b(aU*V) - \lambda [ab](U*V)
    \label{qsh}
\eeq
$\lambda \in \K$ being an arbitrary parameter. The ''merged'' term
$[ab] \in \A$ gives a letter. The empty word
$1 \in \T(\A)$ gives the product unit, i.e. $1*U=U*1=U$, $U \in \T(\A)$.\\
In \cite{LG1} Guo {\emph{et al.}} gave a non-recursive definition
of the quasi-shuffle product by introducing the notion of mixed
shuffles (see later in the section 4).\\
For the parameter $\lambda$ being zero we get the ordinary shuffle product.\\
The augmented quasi-shuffle product on the augmented tensor
module:
\beq
    {\bar{\T}}(\A) :=  \bigoplus_{n>0} \A^{{\otimes}^n}
\eeq
is defined in the following way:
\beq
    aU \boxdot bV := [ab](U*V)
    \label{mqsh}
\eeq
whereby the unit of this product is given by the algebra unit $e \in \A$,
$e \boxdot U = U \boxdot e = U$, $U \in {\bar{\T}}(\A)$.\\
The associativity and commutativity of $\boxdot$ for arbitrary
$\lambda \in \K$ follows from the associativity and commutativity of the
quasi-shuffle product and the product $[\;]$ in $\A$.\\
We define now a modified quasi-shuffle product on $\T(\A)$ using the
algebra unit $e \in \A$:
\beq
    aU \circledast bV := a(U \circledast bV) + b(aU \circledast V)
                                             - \lambda \:e\:[ab](U \circledast V).
    \label{eqsh}
\eeq
For the empty word we have $1 \circledast U = U \circledast 1 := U$, $U \in \T(\A)$.\\
We prove the associativity property (commutativity being obvious)
for this product by induction on the length of words. The
linearity allows us to reduce the proof to generators in $\T(\A)$.
Choose $a,b,c \in \T(\A)$, $\l(a+b+c)=3$:
\beqn
     a \circledast (b \circledast c) &=& a \circledast \big(bc + cb - \lambda e[bc]\big)
                                         \nonumber\\
                                     &=& abc + b\big(ac + ca - \lambda e[ac]\big) - \lambda e[ab]c
                                         \nonumber\\
                                     & & \;\;\;\;\;\;\;\;\;
                                          + acb + c\big(ab + ba - \lambda e[ab]\big) - \lambda e[ac]b
                                         \nonumber\\
                                     & & \;\;\;\;\;\;\;\;\;
                                         - \lambda \big(ae[bc] + e(a[bc] + [bc]a - \lambda e[a[bc]])\big)
                                         +\lambda^2 (e[ae][bc])
                                         \nonumber\\
                                     &=& abc + bac + bca + acb + cab + cba
                                         \nonumber\\
                                     & & - \lambda\big(be[ac]+ e[ab]c+ ce[ab]+ e[ac]b
                                                   + ae[bc]+ea[bc]+e[bc]a\big)
                                         \nonumber\\
                                     & & \;\;\;\;\;\;\;\;\;
                                         + \lambda^2 \big(ee[a[bc]] + ea[bc]\big)
                                         \nonumber
\eeqn
\beqn
   (a \circledast b) \circledast  c  &=& (ab + ba - \lambda e[ab]) \circledast c
                                         \nonumber\\
                                     &=& a\big(bc + cb - \lambda e[bc]\big) + cab -\lambda e[ac]b
                                         \nonumber\\
                                     & & \;\;\;\;\;\;\;\;\
                                          + b\big(ac + ca -\lambda e[ac]\big) + cba - \lambda e[bc]a
                                         \nonumber\\
                                     & & \;\;\;\;\;\;\;\;\
                                        - \lambda \big(e([ab]c + c[ab] - \lambda e[[ab]c]) + ce[ab]\big)
                                         + \lambda^2 (e[ec][ab])
                                         \nonumber\\
                                     &=& abc + acb + cab + bac + bca + cba
                                         \nonumber\\
                                     & & \;
                                         - \lambda\big(ae[bc] + e[ac]b + be[ac] + e[bc]a
                                                   + e[ab]c + ec[ab] + ce[ab]\big)
                                         \nonumber\\
                                     & & \;\;\;\;\;\;\;\;\
                                         + \lambda^2 \big(ee[[ab]c] + ec[ab]\big)
                                         \nonumber
\eeqn
We see already here that this product is associative if and only
if $\lambda = 1$, eliminating the two unwanted terms in the
$\lambda^2$-part, i.e. $ec[ab]$ and $ea[bc]$!
Now we assume that associativity is true for $\l(U+V+W)=k-1>3$ and
choose $aX, bY, cZ \in \T(\A)$, $\l(aX+bY+cZ)=k$ (we use now
$\lambda = 1$):
\beqn
     aX \circledast (bY \circledast cZ) &=& aX \circledast
                                         \big(b(Y \circledast cZ) + c(bY \circledast Z)
                                          - e[bc](Y \circledast Z) \big)
                                         \nonumber\\
                                     &=& aX \circledast b(Y \circledast cZ)
                                         + aX \circledast c(bY \circledast Z)
                                         \nonumber\\
                                     & & \;\;\;\;\;\;\;\;\;\;\;\;\;\;\;\;\;\;\;\;\;\;\;\;\;\;\;\;\;\;\;\;\;
                                         - aX \circledast e[bc](Y \circledast Z)
                                         \nonumber\\
                                     &=&  a\big(X \circledast b(Y \circledast cZ)\big)
                                          + b\big(aX \circledast (Y \circledast cZ)\big)
                                          \nonumber\\
                                     & & \;\;\;\;\;\;\;\;\;\;\;\;\;\;\;\;\;\;\;\;\;\;\;\;\;\;\;\;\;\;\;\;\;
                                          - e[ab]\big(X \circledast (Y \circledast cZ)\big)
                                           \nonumber\\
                                     & & + a\big(X \circledast c(bY \circledast Z)\big)
                                          + c\big(aX \circledast (bY \circledast Z)\big)
                                          \nonumber\\
                                     & & \;\;\;\;\;\;\;\;\;\;\;\;\;\;\;\;\;\;\;\;\;\;\;\;\;\;\;\;\;\;\;\;\;
                                          -  e[ac]\big(X \circledast (bY \circledast Z)\big)
                                           \nonumber\\
                                     & & - a\big(X \circledast e[bc](Y \circledast Z)\big)
                                          + e\big(aX \circledast [bc] (Y \circledast Z)\big)
                                          \nonumber\\
                                     & & \;\;\;\;\;\;\;\;\;\;\;\;\;\;\;\;\;\;\;\;\;\;\;\;\;\;\;\;\;\;\;\;\;
                                          + e[ae]\big(X \circledast [bc](Y \circledast Z)\big)
                                           \nonumber\\
                                     &=&  a\big((X \circledast bY) \circledast cZ\big) +
                                         \;\;        b\big((aX \circledast Y) \circledast cZ\big) -
                                           \nonumber\\
                                     & & c\big((aX \circledast bY) \circledast Z\big)
                                         \;\;\;       + ee[[ab]c]\big((X \circledast Y) \circledast Z\big) -
                                           \nonumber\\
                                     & & e[ac]\big((X \circledast bY) \circledast Z\big)
                                         \;\;\;\;   - e[bc]\big((aX \circledast Y) \circledast Z\big)
                                         \nonumber\\
                                     & & \;\;\;\;\;       -  e[ab]\big((X \circledast Y) \circledast cZ\big)
                                         \nonumber\\
                                     &=& (aX \circledast bY) \circledast cZ
                                         \nonumber\\
                                     & & \hspace{8cm}   \blacksquare \nonumber
\eeqn
The modified quasi-shuffle is homogeneous in contrast to the
ordinary quasi-shuffle product (\ref{qsh}). Anticipating the
result in the next section we mention here that this fact reflects
the homogeneity difference between the Rota-Baxter relation
(\ref{RBR}) and equation (\ref{ANR}).\\
The augmented modified quasi-shuffle product on the augmented
tensor module ${\bar{\T}}(\A)$ is defined as follows:
\beq
    aU \boxtimes bV := [ab](U \circledast V)
    \label{meqsh}
\eeq
\beq
    e \boxtimes V = V \boxtimes e = V.
\eeq
We define now the following two linear operators
\makebox{$B^+_e,\; B^-: {\bar{\T}}(\A) \to {\bar{\T}}(\A)$}
on the augmented tensor module ${\bar{\T}}(\A)$:
\beqn
    B^+_e(a_1 \dots a_n) &:=& ea_1 \dots a_n   \label{plusop}\\
    B^-(a_1 \dots a_n)   &:=&  a_2 \dots a_n.   \label{minusop}
\eeqn
The first one is later to be identified as the Nijenhuis operator
with respect to the augmented modified quasi-shuffle product
(\ref{meqsh}), i.e. in the following section we will show that the
triple \makebox{$({\bar{\T}}(\A), \boxtimes, B^+_e)$} defines a
Nijenhuis algebra, moreover we will see that it fulfills the
universal property.
%
%
%
%
%
\section{Free Associative Nijenhuis Algebra}
%
We will use the augmented modified quasi-shuffle (\ref{meqsh})
product to give the construction of the free Nijenhuis
$\K$-algebra. The existence of such an object follows from general
arguments in the theory of universal algebras \cite{Rota4, PC}
since the category of Nijenhuis $\K$-algebras, defined through the
identity (\ref{ANR}) forms a variety (in the sense of universal
algebras, see especially section 6 of \cite{Rota4} for a concise
summary of the main ideas).\\
Let us remark here that the construction of the free Rota-Baxter
algebra (of weight $\lambda \ne 0$) uses the augmented
quasi-shuffle product (\ref{mqsh}) and works analogously to what
follows. The quasi-shuffle essentially embodies the structure of
relation (\ref{RBR}). The case $\lambda=0$ gives the ''trivial''
Rota-Baxter algebra, i.e. relation (\ref{RBR}) without the second
term on the lefthand side. This construction was essentially given
in \cite{LG1} using a non-recursive notion of a so called mixed
shuffle product. However the formulation using Hoffman's
quasi-shuffle is new. The essential point (for both constructions
of the free objects) lies in the (augmented) quasi-shuffle product
(or mixed shuffle in \cite{LG1}) and its ''merging'' of letters
$\lambda [ab],\; a,b \in \A$ in the third term on the righthand
side of (\ref{mqsh}), whereby $\lambda$ is arbitrary. It
represents the second term on the lefthand side of (\ref{RBR}).
The modified quasi-shuffle product (\ref{meqsh}) reflects relation
(\ref{ANR}), in particular, the third term on the righthand side
of it is related to the second term on the lefthand side of the
associative Nijenhuis relation. It has to be underlined that the
associativity of (\ref{meqsh}) relies on the minus sign on the
righthand side. This is due to the fact \cite{CGM} that the
following bilinear map on a $\K$-algebra $\A$, $N$ being an
arbitrary $\K$-linear map:
\beq
    \mu_{N}(a,b):=N(a)b + aN(b) - N(ab)
    \label{Nprod}
\eeq
gives an associative product if and only if the following map,
called $\mu$-Nijenhuis torsion in \cite{CGM}:
\beq
    T_{\mu,N}(a,b) := N(\mu_N(a,b)) - N(a)N(b)
    \label{NT}
\eeq
is a 2-Hochschild cocycle. This is especially true for $N$ being a
Nijenhuis operator holding exactly relation (\ref{ANR}).\\
We now show that $({\bar{\T}}(\A), \boxtimes, B^+_e)$ is a
Nijenhuis algebra, i.e. the operator $B^+_e$ holds the associative
Nijenhuis relation:
\beqn
    \B(U) \bp \B(V) + (\B)^2(U \bp V) &=& [ee](U \circledast V) +
                                          \nonumber\\
                                      & & \;\;\;\;\;\;\;\;
                                          ee\big([u_1v_1](B^-(U) \circledast B^-(V))\big)
                        \nonumber\\
                                      &=& e\big(u_1(B^-(U) \circledast V)
                                                               + v_1(U \circledast B^-(V))
                        \nonumber\\
                                      & & \;\;\;\;\;
                                                   -e[u_1v_1](B^-(U) \circledast B^-(V))\:\big)
                        \nonumber\\
                                      & & \;\;\;\;\;\;
                                          \;\;\; + ee[u_1v_1]\big(B^-(U) \circledast B^-(V)\big)
                        \nonumber\\
                                      &=& \B\big(U \bp \B(V) + \B(U) \bp V\big)\\
                                      & & \hspace{5.5cm}   \blacksquare
                        \nonumber
\eeqn
For the triple $({\bar{\T}}(\A), \boxtimes, B^+_e)$ to be the free
Nijenhuis algebra over $\A$ we have to show that the universal
property holds. Let $\U$ be an arbitrary Nijenhuis $\K$-algebra
with Nijenhuis operator $N$ and $\phi: \A \to \U$ an
$\K$-algebra homomorphism.\\
We have to extend the $\K$-algebra map $\phi$ to a Nijenhuis homomorphism
(\ref{NP}) \makebox{${\tilde{\phi}}: \aT \to \U$} using the following
important fact: every generator $a_1 \dots a_n \in \aT$ can be
written using the modified augmented quasi-shuffle product:
\beq
    a_1 \dots a_n = a_1 \bp \B \big(a_2 \bp \B(a_3 \bp \B( \dots B\big(a_{n-1} \bp \B(a_n)\big) \dots ))\big)
    \label{fact}
\eeq
Since the extension ${\tilde{\phi}}$ is supposed to be a Nijenhuis
algebra homomorphism we have the unique possible extension:
\beq
    {\tilde{\phi}}(a_1 \dots a_n):= \phi(a_1)
                                      N\big(\phi(a_2)N( \dots N\big(\phi(a_{n-1}) N(\phi(a_n))\big) \dots) \big).
    \label{uprop}
\eeq
Defining the map \makebox{$N_w : \U \to \U,\; w \in \U$,
$N_w(x):=N(wx),\; x \in \U$} we can write (\ref{uprop}) like
\cite{LG1}:
\beq
    {\tilde{\phi}}(a_1 \dots a_n) = \phi(a_1)\: \big\{ \circ^{n}_{i=2}N_{\phi(a_i)}(1_{\U}) \big\}.
\eeq
So we are left to show that ${\tilde{\phi}}$ is a Nijenhuis
homomorphism. By construction it is a well defined $\K$-linear
map.
The Nijenhuis property (\ref{NP}) follows by \cite{LG1}:
\beqn
    {\tilde{\phi}}(\B(a_1 \dots a_n)) &=&  {\tilde{\phi}}(ea_1 \dots a_n)                                \nonumber\\
                                      &\stackrel{(\ref{uprop})}{=}&  \circ^{n}_{i=1}N_{\phi(a_i)}(1_{\U})\nonumber\\
                                      &=&  N \big(\phi(a_1) \circ^{n}_{i=2}N_{\phi(a_i)}(1_{\U})\big)    \nonumber\\
                                      &=&  N \big({\tilde{\phi}}(a_1 \dots
                                      a_n)\big). \label{trick}\\
                                      & & \hspace{8cm}   \blacksquare                                    \nonumber
\eeqn
Finally we would like to show that $\tilde{\phi}$ is a
$\K$-algebra homomorphism, i.e. preserves multiplication. This we
proof by induction on the length of words,  $\l(X+Y)=m+n=k$. For
$k=2$ we have, $a,b \in \aT$:
\beqn
    \ta(a \bp b) &=& \ta([ab])                                  \nonumber\\
                 &\stackrel{(\ref{uprop})}{=}& \phi([ab])       \nonumber\\
                 &=& \phi(a)\phi(b) = \ta(a)\ta(b).             \nonumber
\eeqn
Let $X:=u_1 \dots u_n \in \A^{\otimes^{n}},\; Y:=v_1 \dots v_m \in
\A^{\otimes^{m}}$, of length $m+n>2$:
\beqn
    \ta(X \bp Y) &=& \ta \big( [u_1 v_1] \: \{B^-(X) \circledast B^-(Y)\}\:\big)
                     \nonumber\\
                 &=& \ta \big( [u_1 v_1] \: \{\B(B^-(X)) \bp B^-(Y)  +  B^-(X) \bp \B(B^-(Y))
                     \nonumber\\
                 & & \;\;\;\;\;\;\;\;\;\;\;\;\;\;\;\;\;\;\;\;- \B(B^-(X) \bp B^-(Y))\: \}\big)
                     \nonumber\\
                 &\stackrel{(\ref{uprop})}{=}& \phi(u_1)\phi(v_1) \: N\big(\: \ta \{ \: \B(B^-(X))\bp B^-(Y)
                     \nonumber\\
                 & & + B^-(X)\bp \B(B^-(Y)) - \B(B^-(X) \bp B^-(Y))\: \}\:\big)
                     \nonumber\\
                 &=& \phi(u_1)\phi(v_1)\: N\big( \: \ta(\B(B^-(X))) \: \ta(B^-(Y))
                     \nonumber\\
                 & & + \ta(B^-(X)) \: \ta(\B(B^-(Y))) - N\{\ta(B^-(X)) \: \ta(B^-(Y))\}\:\big)
                     \nonumber\\
                 &=& \phi(u_1)\phi(v_1) \: N\big\{ \: N\big(\ta(B^-(X))\big) \: \ta(B^-(Y))
                     \nonumber\\
                 & &  + \ta(B^-(X)) \: N\big(\ta(B^-(Y))\big) - N\big(\ta(B^-(X)) \: \ta(B^-(Y)\big) \: \big\}
                     \nonumber\\
                 &\stackrel{(\#)}{=}& \phi(u_1)\phi(v_1)\: N\{\ta(B^-(X))\} \: N\{\ta(B^-(Y))\}
                     \nonumber\\
                 &=& \phi(u_1) N\{\ta(B^-(X))\} \; \phi(v_1) \: N\{\ta(B^-(Y)\}
                     \nonumber\\
                 &=& \ta(X) \: \ta(Y)
                     \nonumber\\
                 & & \hspace{9.6cm}   \blacksquare
                     \nonumber
\eeqn
In the third line we used the same trick as in (\ref{trick}).
Apparently by going from line $(\#)$ to the next we see that the
use of the augmented modified quasi-shuffle product in the above
construction of the
free Nijenhuis algebra is limited to the commutative case.\\
Using the augmented quasi-shuffle product (\ref{mqsh}) the proof
of the universal property for the triple  $({\bar{\T}}(\A),
\boxdot, \lambda, B^+_e)$ goes analogously, giving the free
Rota-Baxter $\K$-algebra of weight $\lambda$ over $\A$.\\
The algebras $(\T(\A), *, \lambda, B^+_e)$ and $(\T(\A),
\circledast, B^+_e)$ define a Rota-Baxter algebra and a Nijenhuis
algebra, respectively.\\
The above construction may also be used to give the free Nijenhuis
algebra on a set $S$ by working with the polynomial algebra
$\K[S]$.\\[0.4cm]
%
%
%
%
%
{\emph{Mixed Shuffle Product}}

Guo's {\emph{et al}} mixed shuffle product in \cite{LG1} may be
described the following way. Let $Z:=z_1 \dots z_n,\; X:=x_{n+1}
\dots x_{n+m} \in \aT$ be two words. The mixed shuffle relies on
so called admissible pairs $(k,k+1),\; k\in \N$, defined as
follows. First take the ordinary shuffle product of $X,\:Z \in
\aT$, $\l(X+Z)= m+n$, i.e. relation (\ref{qsh}) with $\lambda=0$:
\beqn
  X * Z &:=& x_1(B^-(X)*Z) + z_1(X*B^-(Z))            \nonumber\\
        &=:& \sum _{\tau = 1}^{\Gamma_{m,n}} W_{\tau}^{(X,Z)}
        \;\;\; \in \A^{\otimes^{m+n}}.
             \label{shSum}
\eeqn
$\Gamma_{m,n}:=\binom{m+n}{m}$. Every pair of consecutive letters
in one of the words $W_{\tau}^{(X,Z)}$ in the above shuffle sum
(\ref{shSum}) of the form $\dots x_iz_j \dots$, $x_i$ being the
$k$-th letter in this word gives an admissible pair $(k,k+1)$. We
denote the set of admissible pairs by
\makebox{$P_{W_{\tau}^{(X,Z)}}:=\{(k,k+1)|\: 1 \le k < m+n \}$}. A
word may contain no, one or several admissible pairs. If it
contains any there is no change and we just keep the word
$W_{\tau}^{(X,Z)}$. If it contains admissible pairs, we add the
following sum to the above shuffle product (\ref{shSum}):
\beq
    \sum_{p \subset P_{W_{\tau}^{(X,Z)}}} (-\lambda)^{|p|} W^{[(X,Z)]_p}_{\tau}
     \;\;\; \in \bigoplus_{i=1}^{m+n-1}\A^{\otimes^{i}}.
     \label{msh}
\eeq
Keeping the word $W_{\tau}^{(X,Z)}$ in (\ref{shSum}), the empty
set must be excluded \makebox{$p \ne \varnothing$} in (\ref{msh}). The word
$W_{\tau}^{[(X,Z)]_p}$ denotes the original (''shuffled'') string
but with consecutive letters at position \makebox{$k,\; (k,k+1) \in p
\subset P_{W_{\tau}^{(X,Z)}}$} being ''merged'' (in $\A$) to
one letter, i.e. $\dots [x_iz_j] \dots $.\\[0.4cm]
%
%
%
%
%
{\emph{Dendriform Di- and Trialgebra Structures}}

We mentioned in the beginning the relation between 
Rota-Baxter algebras and Loday-type algebras \cite{KEF, Agu1, Agu2},
i.e. dendriform di- and trialgebra structures (see \cite{JLL1, JLLR} for
definitions).\\
In this final part we would like to relate the above given constructions
of the free 
Rota-Baxter algebra, i.e. especially the
augmented and augmented modified quasi-shuffle products to these Loday-type
structures.\\
In \cite{R} Ronco defined a dendriform dialgebra structure on the
tensor coalgebra ${\tilde{\T}}(V)$ over a vector space $V$ in
connection to the ordinary shuffle product (\ref{shSum}). The two
compositions $\prec,\; \succ $ are defined using the fact, that
the shuffle sum (\ref{shSum}) may be characterized by words either
beginning with $x_1$ or $z_1$:
\beqn
     X \prec Z &:=& x_1(B^-(X) * Z) \nonumber\\
     X \succ Z &:=& z_1(X * B^-(Z)) \nonumber.
\eeqn
In the context of $\aT$ over the $\K$-algebra $\A$ we see how this relates
to the augmented shuffle product (\ref{mqsh}), respectively the ''trivial''
Rota-Baxter map $\B$ of weight $\lambda=0$:
\beqn
     X \prec Z &:=& X \boxdot \B(Z) \nonumber\\
     X \succ Z &:=& \B(X) \boxdot Z \nonumber,
     \label{RBD}
\eeqn
which gives the dendriform dialgebra structure related to the
Rota-Baxter relation of weight zero found by Aguiar \cite{Agu1}.
Using \makebox{$\B$ and $\lambda id_{\aT}-\B=:{\widehat{\B}}$} the
dendriform compositions \makebox{$X \prec Z:= X \boxdot \B(Z),\; X
\succ Z:=-{\widehat{\B}}(X) \boxdot Z$} relate to the augmented
quasi-shuffle product, i.e. to the Rota-Baxter relation of weight
$\lambda \ne 0$ \cite{KEF}.\\
Loday and Ronco related in \cite{JLLR} the dendriform trialgebra
structure to Hoffman's quasi-shuffle product:
\beqn
     X \prec Z &:=& x_1(B^-(X) * Z) = X \boxdot \B(Z)
     \label{T1}\\
     X \succ Z &:=& z_1 (X * B^-(Z)) = \B(X) \boxdot Z
     \label{T2}\\
     X \bullet Z &:=& [x_1z_1]\big(B^-(X) * B^-(Z)\big) = X \boxdot Z.
     \label{T3}
\eeqn
Including the $\lambda$-term in the $\boxdot$-product in
(\ref{RBD}) we might describe this structure using the free
Rota-Baxter algebra  \makebox{$({\bar{\T}}(\A), \boxdot, \lambda,
B^+_e)$} of weight $\lambda \ne 0$. The last term is just the
product $X \boxdot Z$ (\ref{mqsh}). 
Let us remark here an observation with respect to the work
\cite{R} of Ronco. From the previous results it follows that
Rota-Baxter 
algebras always give dendriform algebra
structures. We used in the definition of the extension of an
algebra homomorphism to a 
Rota-Baxter
homomorphism the fact
(\ref{fact}), that the elements in $\aT$ may be written using the
%
%
Rota-Baxter map $\B$ ($\boxdot$ instead of $\boxtimes$):
\beqn
    a_1 \dots a_n &=& a_1 \bd \B \big(a_2 \bd \B(a_3 \bd \B( \dots \B\big(a_{n-1} \bd
                                                          \B(a_n)\big) \dots))\big)
                      \nonumber\\
                  &=& \B \big( \dots \B \big(\B \big (\B (a_n) \bd a_{n-1}\big)
                       \bd a_{n-2}\big) \bd a_{n-3} \dots \big) \bd a_1
                       \nonumber.
\eeqn
With respect to the dendriform (trialgebra) structures this may be
written using either of the compositions $\prec,\; \succ $ and
introducing the ''identity'' maps $\omega_{\prec, \succ}$ on $\aT$
($a_i \in \A$ being letters):
\beqn
    \omega_{\prec}(a_1 \dots a_n) &=& a_1 \prec \big(a_{2} \prec (a_{3} \prec
                  ( \dots \prec \big(a_{n-1} \prec a_n \big) \dots ))\big)
\eeqn
These maps are of importance with respect to the primitive
elements and brace-algebras \cite{R, R2} in a dendriform Hopf
algebra. The connection to Loday's recent work \cite{JLL3} has to
be clarified in a future work.
%
%
%
%
%
\section{Summary and Outlook}
%

%
We have shown that Hoffman's (augmented) quasi-shuffle product
essentially gives the free commutative associative unital
Rota-Baxter algebra on a commutative associative unital
$\K$-algebra. As the associative Nijenhuis relation may be
regarded as the homogeneous version of the Rota-Baxter relation
the natural way to construct a free commutative associative
Nijenhuis algebra is to ''homogenize'' the (augmented)
quasi-shuffle product. Commutativity following by construction, we
showed explicitly the associativity. Due to the homogeneity of the
associative Nijenhuis relation the sign, i.e. free parameter, in
front of the ''merging'' part in the modified quasi-shuffle
product has to be fixed to be $1$. We then showed that this new
product effectively gives the free associative Nijenhuis structure
generated by a commutative associative unital $\K$-algebra. The
''shift'' operator $\B$ acts in the former case as the Rota-Baxter
map and as the Nijenhuis operator in the latter.\\
Finally we related these free objects to the fascinating
dendriform di- and trialgebra structures given in the work of
Loday and Ronco. Further work concerning this link and the
algebraic properties of the augmented modified quasi-shuffle
product has to be done. Following the work of Guo {\emph{et al.}}
on the free Rota-Baxter algebra further investigations into the
structure of associative Nijenhuis algebras need to be done.\\
Both the Rota-Baxter and the associative Nijenhuis relation are of
interest with respect to the Hopf algebraic formulation of the
theory of renormalization in perturbative Quantum Field Theory and
especially to aspects of integrable systems.\\[0.5cm]
\noindent
{\emph{Acknowledgments}}:

A special vote of thanks goes to Prof. D. Kreimer for helpful
comments concerning this work. This work was supported in parts by NSF grant 
DMS-0205977 at the Center for Mathematical Physics at Boston University. 
I am also indebt to Prof. Loday for his remarks. 
The author is supported by the German Academic Exchange Service (DAAD).  
%
%
%

%
%
%
%
%

\begin{thebibliography}{AB}
%
%
\normalsize
%
%
%
%
%
%
 \bibitem{CGM} J. Cari{\~{n}}ena, J. Grabowski, G. Marmo,
 {\emph{''Quantum bi-Hamiltonian systems''}},
 Internat. J. Modern Phys. A, {\bf{15}}, 2000, no. 30, 4797--4810.
%
%
%
%
%
 \bibitem{Rota1} G.-C. Rota,
  {\emph{''Ten mathematics problems I will never solve''}},
  Mitt. Dtsch. Math.-Ver. 1998, no. {\bf{2}}, 45--52.
%
 \bibitem{Rota2} G.-C. Rota,
  {\emph{''Baxter operators, an introduction''}}, in ''Gian-Carlo Rota on combinatorics:
  Introductory papers and commentaries'', J. P.S. Kung Ed.,
  Contemp. Mathematicians, Birkh\"auser Boston, Boston, MA, 1995, 504--512.
%
 \bibitem{Rota3} G.-C. Rota,
  {\emph{''Baxter algebras and combinatorial identities. I, II.''}}, Bull. Amer. Math. Soc. {\bf{75}}, 1969,
  325-329; ibid. {\bf{75}}, 1969, 330--334.
%
 \bibitem{Rota4} G.-C. Rota, D. Smith,
  {\emph{''Fluctuation theory and Baxter algebras''}}, in ''Gian-Carlo Rota on combinatorics:
  Introductory papers and commentaries'', J. P.S. Kung Ed.,
  Contemp. Mathematicians, Birkh\"auser Boston, Boston, MA, 1995, 481--503.
%
 \bibitem{A} F. V. Atkinson,
  {\emph{''Some aspects of Baxter's functional equation''}},
  J. Math. Anal. Appl. {\bf{7}}, 1963, 1--30.
%
 \bibitem{B} G. Baxter,
  {\emph{''An analytic problem whose solution follows from a simple algebraic identity''}},
  Pacific J. Math. {\bf{10}}, 1960, 731--742.
%
 \bibitem{C} P. Cartier,
  {\emph{''On the structure of free Baxter algebras''}},
  Advances in Math. {\bf{9}}, 1972, 253--265.
%
 \bibitem{KEF} K. Ebrahimi-Fard,
  {\emph{''Loday-type algebras and the Rota-Baxter relation''}},
  Letters in Mathematical Physics, {\bf{61}},  no. 2, 2002, 139--147.
%
%
%
%
%
 \bibitem{GS1} I.Z. Golubchik, V.V. Sokolov,
  {\emph{''One more type of classical Yang-Baxter equation''}},
  Funct. Anal. Appl., {\bf{34}}, no. 4, 2000, 296--298
%
 \bibitem{GS2} I.Z. Golubchik, V.V. Sokolov,
 {\emph{''Generalized Operator Yang-Baxter Equations, Integrable ODEs and Nonassociative
 Algebras''}}, J. of Nonlinear Math. Phys., {\bf{7}}, No.2, 2000, 184--197
%
 \bibitem{BD} Belavin, Drinfeld,
  {\emph{''Triangle Equations and Simple Lie-Algebras''}},
  Classic Reviews in Mathematics and Mathematical Physics, 1. Harwood Academic Publishers,
  Amsterdam, 1998, viii+91 pp.
%
 \bibitem{KSM} Y. Kosmann-Schwarzbach, F.Magri,
  {\emph{''Poisson-Nijenhuis structures''}},
  Ann. Inst. H. Poincaré Phys. Théor., {\bf{53}}, no. 1, 1990, 35--81.
%
 \bibitem{KS} Y. Kosmann-Schwarzbach,
 {\emph{''The modified Yang-Baxter equation and bi-Hamiltonian structures''}},
 Differential geometric methods in theoretical physics (Chester, 1988), 12--25,
 World Sci. Publishing, Teaneck, NJ, 1989.
%
%
%
%
%
 \bibitem{Kreimer1} D. Kreimer,
  {\emph{''Structures in Feynman Graphs - Hopf Algebras and Symmetries''}},
  BUCMP-02-02, Feb 2002, preprint: hep-th/0202110.
%
 \bibitem{Kreimer2} D. Kreimer,
  {\emph{''New Mathematical Structures in Renormalizable Quantum Field Theory''}},
  Annals Phys., {\bf{303}}, 2003, 179--202.
%
%
%
%
%
 \bibitem{LG1} L. Guo, W. Keigher,
 {\emph{''Baxter algebras and shuffle products''}},
 Adv. Math., {\bf{150}}, no. 1, 2000, 117--149.
%
 \bibitem{LG2} L. Guo,
 {\emph{''Baxter algebras and differential algebras''}}, in ''Differential algebra and related topics'',
 (Newark, NJ, 2000), 281--305, World Sci. Publishing, River Edge, NJ, 2002
%
%
%
%
%
 \bibitem{Agu1} M. Aguiar,
 {\emph{''Prepoisson algebras''}},
 Letters in Mathematical Physics, {\bf{54}}, no. 4, 2000, 263--277.
%
 \bibitem{Agu2} M. Aguiar,
 {\emph{''Infinitesimal bialgebras, pre-Lie and dendriform algebras''}},
 Nov. 2002, preprint: math.QA/0211074, to appear in ''Hopf
 Algebras: Proc. from an Int. Conf. held at DePaul University'',
 Marcel Dekker.
%
%
%
%
%
 \bibitem{JLL1} J.-L. Loday,
  {\emph{''Dialgebras''}}, in ''Dialgebras and related operads''
  Springer Lecture Notes in Mathematics {\bf{1763}}, 2001, 7--66.
%
 \bibitem{JLLR} J.-L. Loday, M. Ronco,
  {\emph{''Trialgebras and families of polytopes''}}, Mai 2002, preprint: math.AT/0205043
%
 \bibitem{JLL3} J.-L. Loday,
  {\emph{''Scindage d'associativit{\'e} et alg{\`e}bre de Hopf''}},
  Jan. 2003, preprint on http://www-irma.u-strasbg.fr/~loday.
%
 \bibitem{R} M. Ronco,
  {\emph{''Primitive elements in a free dendriform algebra''}},
  New trends in Hopf algebra theory (La Falda, 1999),
  Contemp. Math., {\bf{267}}, 2000, 245--263.
%
 \bibitem{R2} M. Ronco,
  {\emph{''Eulerian idempotents and Milnor-Moore theorem for certain non-commutative Hopf algebras''}},
  J. Algebra, {\bf{10}}, 2002, 152--172.
%
%
%
%
 \bibitem{MH} M. Hoffman,
 {\emph{''Quasi-shuffle products''}},
 J. Algebraic Combin., {\bf{11}}, no. 1, 2000, 49--68.
%
%
%
 \bibitem{PC} P. Cohn,
 {\emph{''Universal Algebra''}}, Second edition. Mathematics and its Applications,
 {\bf{6}}., D. Reidel Publishing Co., Dordrecht-Boston, Mass., 1981. xv+412 pp.
%
%
%
%
\end{thebibliography}
\end{document}